\documentstyle{article}  
\title{ {\LARGE {\bf  Supersymmetric WZW $\sigma$ Model on Full and Half
Plane } }}
\author{ {\Large R. A. Zait and M. F. Mourad}  \\[16pt] 
{\it Mathematics Department, Faculty of Science, Minia University, 
Minia, Egypt.} } 
\date{}
\begin{document}
\maketitle
\topmargin 0cm
\oddsidemargin 0pt
\evensidemargin 0pt
{\Large
\vfill
\centerline{{\bf Abstract}}
\centerline{}
We study classical integrability of the supersymmetric $U(N)$ $\sigma$ model 
with the Wess-Zumino-Witten term on full and half plane.
We demonstrate the existence of nonlocal conserved currents of the model and
derive general recursion relations for the infinite number of the 
corresponding charges in a superfield framework. 
The explicit form of the first few supersymmetric charges are constructed.
We show that the considered model is integrable on full plane as a concequence 
of the conservation of the supersymmetric charges. 
Also, we study the model on half plane with free boundary, and examine 
the conservation of the supersymmetric charges
on half plane and find that they are conserved as a result of the equations
of motion and the free boundary condition. As a result, the 
model on half plane with free boundary is integrable. Finally, we conclude
the paper and some features and comments are presented.\\[20pt]
PACS numbers:\quad 11.10.Kk\qquad 11.10.Ef\qquad 11.25.-w\qquad 11.30.Pb \\[20pt]
Keywords:\quad $U(N)$ sigma model $-$ Wess-Zumino-Witten term $-$
Supersymmetry  $-$ Boundary condition $-$ Integrability.

\vfill
\setcounter{equation}{0}
\section{Introduction}
Quantum field theory provides a powerful and unifying language to understand
a variety of physical phenomena. In recent years, nonlinear $\sigma$ models
in two-dimensional Euclidean space have become an increasingly important area
of research \cite{1}. There are various reasons for this: the models appear
to be the low-dimensional analogues of four-dimensional gauge theories,
they often arise as approximate models in the contexts of both particle
physics and solid state physics, they provide examples of integrable systems,
..., etc. Motivated by some observations from the geometric properties of the
strings, the interest has shifted towards $\sigma$ models with the 
Wess-Zumino-Witten (WZW) term \cite{2}. This term was first introduced by
Wess and Zumino \cite{3} who, in the context of the chiral theory of pions,
showed that it represents the effects of flavour anomalies. It was 
reintroduced by Polyakov and Wiegmann \cite{4}, and Witten \cite{5}, who
studied various effects associated with its presence in different models.
Moreover, Witten \cite{2} has pointed out that as this term breaks some
reflection symmetries of the manifold, it should be included in any low energy
approximation to QCD based on Skyrme ideas \cite{5} in which the original
Skyrme model possesses too many symmetries. 
The more physical models should include fermions. There are many ways of
including fermions into $\sigma$ models, of which the most interesting one is
that which renders the theory supersymmetric (Susy) \cite{6,7}. In the present
paper, we shall study the integrability of the Susy $\sigma$ model with the
WZW-term, which from now we shall call Susy WZW $\sigma$ model, on both full
and half plane.
\par
Field theories with boundaries have been attracting the attention of 
theoretical physicists for a number of reasons \cite{8}. For example, the
existence of boundaries in the study of string theory can be used to
distinguish different types of string theories and are also regarded as the
reason for the occurrence of noncommutativity on the $D$-branes. The study
of such field theories is not only intrinsically interesting but also provides
a better understanding of boundary related phenomena in statistical physics
and condensed matter \cite{9}. An integrable field theory possesses an 
infinite set of integrals of motion. In the full plane theory, these integrals
of motion follow from an infinite number of divergenceless currents. On the
other hand, when the theory is restricted to a half plane, the existence of
conserved charges on the full plane does not guarantee integrability unless
special boundary conditions are specified.
\par
The classical integrability of the $O(N)$ nonlinear $\sigma$ model on a 
half-line was examined by Corrigan and Sheng \cite{10}, where they established
the existence of an infinity of conserved charges in involution for the free
boundary condition. Moriconi and De Martino \cite{11} have studied the
quantum integrability of the $O(N)$ nonlinear $\sigma$ model and the $O(N)$
Gross-Neveu model on the half-line and found that the first model is 
integrable with Neumann, Dirichlet and a mixed boundary conditions and that 
the latter model is integrable under a certain condition. The problem of
consistent Hamiltonian structure for the $O(N)$ nonlinear $\sigma$ model in
the presence of five different types of boundary conditions is considered in
\cite{12}. Conserved and commuting charges are investigated in both bosonic
and Susy classical chiral models with and without WZW-term by Evans 
{\it et al.} \cite{13}. Also, the $N=1$ Susy two-dimensional nonlinear
$\sigma$ model with boundaries has been considered in \cite{14}.
\par
In this paper, we examine the integrability of the Susy WZW $\sigma$ model as
a concequence of the conservation of the nonlocal charges on both full and 
half plane. This paper is organized as follows: in the following section,
we introduce the Susy WZW $\sigma$ model on full plane. In section 3, we
demonstrate the existence of nonlocal conserved currents of the model and
derive general recursion relations for the infinite number of the 
corresponding charges in a superfield framework. In section 4, we obtain the
explicit form of the first few Susy charges and show that the Susy WZW $\sigma$
model is integrable on full plane as a concequence of the conservation of the
Susy charges. In section 5, we introduce the Susy WZW $\sigma$ model on half
plane with free boundary. We examine the conservation of the Susy charges
on half plane and find that they are conserved as a result of the equations
of motion and the free boundary condition. Concequently, the Susy WZW $\sigma$
model on half plane with free boundary is integrable. Finally, we conclude
the paper and some features and comments
are presented.
\setcounter{equation}{0}
\section{Susy WZW $\sigma$ model on full plane}
In the superspace approach to supersymmetry, a scalar superfield
$\phi(x,y,\theta_1,\theta_2)$ defined on the full two-dimensional superspace
with the coordinates $x,y$ are the standard two-dimensional Euclidean space
and $\theta_1,\theta_2$ are two-components of a real Grassmannian spinor
$\theta$, can be written in the form \cite{7,15}
\begin{equation}
\phi\,=\,g[1\,+\,i\theta_1\psi_2\,-\,i\theta_2\psi_1\,+\,
i\theta_1\theta_2F],
\end{equation}
where $\psi_1(x,y)$ and $\psi_2(x,y)$ are two components of a complex 
anticommuting spinor field $\psi$ (which anticommute with $\theta_1,\theta_2$),
$g(x,y)$ is a bosonic field and $F(x,y)$ is an auxiliary scalar field. We
adopt the conventions of ref. \cite{7}, and introduce the complex variables
$x_\pm =x\pm iy$, $\theta_\pm =\theta_1\pm i\theta_2$, and $\psi_\pm
=\frac{1}{2}(\psi_1\pm i\psi_2)$. As the superfield $\phi$ is a unitary field,
we impose the following constraints on $g,\psi_\pm$ and $F$:
\begin{equation}
g^\dagger g\,=\,1,\qquad \psi_\pm^\dagger\,=\,-\psi_\mp,\qquad
F\,+\,F^\dagger\,=\,2(\psi_+^\dagger\psi_+\,-\,\psi_-^\dagger\psi_-),
\end{equation}
where $\dagger$ denotes the hermitian conjugation.
\par
The action of the Susy WZW $\sigma$ model is given by
\begin{equation}
S\,=\,\int dx\,dy\,d\theta_+ d\theta_-\,L_0
\,+\,ik\int dx\,dy\,dt\,d\theta_+ d\theta_-\,L_1,
\end{equation}
with
\begin{equation}
\begin{array}{ll}
L_0\,=\,\frac{1}{2}\,\hbox{Tr}(D_+\phi^\dagger D_-\phi\,-\,
D_-\phi^\dagger D_+\phi),\\
L_1\,=\,-\hbox{Tr}[(D_+\tilde{\phi}^\dagger D_-\tilde{\phi}\,+\,
D_-\tilde{\phi}^\dagger D_+\tilde{\phi})\tilde{\phi}^\dagger\dot{\tilde{\phi}}
],
\end{array}
\end{equation}
where the supercovariant derivatives $D_\pm =\partial_{\theta_\pm}+i\theta_\pm
\partial_\pm$ with $\partial_{\theta_\pm}=\partial/\partial\theta_\pm$ and
$\partial_\pm=\partial/\partial x_\pm$, and $k$ is a real parameter. The
first term in (2.3) is the Susy $\sigma$ model action of the superfield
$\phi$. The second term in (2.3) is the Susy WZW-term in which $\phi$ has been
extended to $\tilde{\phi}(x,y,t,\theta_1,\theta_2)$ where the additional
variable $t$ satisfies $0\le t\le 1$. Following Witten \cite{5}, we choose
the boundary conditions of this extension to be such that
$\tilde{\phi}=\phi$ at $t=1$, and $\tilde{\phi}=K$ at $t=0$, where $K$ is a
constant unitary matrix.
\par
The equations of motion corresponding to (2.3) can be given by
\begin{equation}
(1-\lambda)D_+(\phi^\dagger D_-\phi)\,-\,
(1+\lambda)D_-(\phi^\dagger D_+\phi)\,=\,0,
\end{equation}
where $\lambda =ik$.
\par
If we define
\begin{equation}
C_\pm\,=\,(1\,\pm\,\lambda)A_\pm\,\qquad \hbox{with}\quad
A_\pm\,=\,\frac{1}{2}\phi^\dagger D_\pm\phi.
\end{equation}
Then the field equations (2.5) become
\begin{equation}
D_+C_-\,-\,D_-C_+\,=\,0.
\end{equation}
Moreover, $C_\pm$ satisfy the zero-curvature condition
\begin{equation}
D_+C_-\,+\,D_-C_+\,+\,2\{C_+\,,\,C_-\}\,=\,0.
\end{equation}
We notice that equations (2.7) and (2.8) can be treated as integrability
conditions of a one-parameter family of linear system.
\par
The field equations (2.7) can be resolved into the components:
\begin{equation}
\partial_+\psi_+\,+\,\frac{1}{2}(1+\lambda)[g^\dagger\partial_+g\,,\,\psi_+]
\,-\,\frac{i}{4}(1+\lambda)(\psi_-F+F^\dagger\psi_-)\,=\,0,
\end{equation}
\begin{equation}
\partial_-\psi_-\,+\,\frac{1}{2}(1-\lambda)[g^\dagger\partial_-g\,,\,\psi_-]
\,+\,\frac{i}{4}(1-\lambda)(\psi_+F+F^\dagger\psi_+)\,=\,0,
\end{equation}
\begin{equation}
(1-\lambda)\partial_+(g^\dagger\partial_-g\,+\,i\psi_+\psi_+)
\,+\,(1+\lambda)\partial_-(g^\dagger\partial_+g\,+\,i\psi_-\psi_-)\,=\,0,
\end{equation}
with $F=(1+\lambda)\psi_+\psi_- -(1-\lambda)\psi_-\psi_+$. Clearly, equation
(2.11) corresponds to the conservation of Noether currents $J_\pm$:
\begin{equation}
\partial_+J_- \,+\,\partial_-J_+\,=\,0,
\end{equation}
where
\begin{equation}
J_\pm\,=\,(1\,\pm\,\lambda)(g^\dagger\partial_\pm g\,+\,i\psi_\mp\psi_\mp)
\,=\,J_{\pm}^B\,+\,J_{\pm}^F,
\end{equation}
with $J_{\pm}^B$ and $J_{\pm}^F$ are the bosonic and fermionic parts of
$J_\pm$.
\par
It is important to note that in terms of the $x-y$ variables, equation (2.12)
can be rewritten as
\begin{equation}
\partial_\mu J_\mu \,=\,0; \quad \mu\,=\,x,\,y
\end{equation}
where
\begin{equation}
J_x\,=\,g^\dagger\partial_xg-i\lambda g^\dagger\partial_yg +
\frac{i}{2}(\psi_1\psi_1-\psi_2\psi_2)+
\frac{\lambda}{2}(\psi_1\psi_2+\psi_2\psi_1)\,=\,
J_x^B\,+\,J_x^F,
\end{equation}
\begin{equation}
J_y\,=\,g^\dagger\partial_yg+i\lambda g^\dagger\partial_xg +
\frac{i}{2}(\psi_1\psi_2+\psi_2\psi_1)-
\frac{\lambda}{2}(\psi_1\psi_1-\psi_2\psi_2)\,=\,
J_y^B\,+\,J_y^F.
\end{equation}
\setcounter{equation}{0}
\section{Nonlocal conserved currents and charges}
Brezin {\it et al.} \cite{16} have discussed the existence of nonlocal 
conserved currents in two-dimensional models. Their discussion is based on
two basic assumptions. Clearly, their assumptions are satisfied in the Susy
WZW $\sigma$ model and are represented by equations (2.5) and (2.6). 
Therefore, we can follow the procedure described in \cite{15,17} and construct
infinite number of nonlocal conserved currents for the Susy WZW $\sigma$
model. We first define the supercovariant derivatives $\tilde{D}_\pm$ as
\begin{equation}
\tilde{D}_\pm\,=\,D_\pm\,+\,2C_\pm.
\end{equation}
Clearly $\tilde{D}_\pm$ satisfy $\{\tilde{D}_+\,,\,\tilde{D}_-\}=0$ and
$\{\tilde{D}_+\,,\,D_-\}=\{\tilde{D}_-\,,\,D_+\}$. The construction of
nonlocal conserved currents is inductive. Let the first currents given by
\begin{equation}
\begin{array}{ll}
V_+^{(1)}\,\equiv\,2C_+\,=\,-D_+\chi^{(1)},\\
V_-^{(1)}\,\equiv\,2C_-\,=\,D_-\chi^{(1)},
\end{array}
\end{equation}
where the Susy charge $\chi^{(1)}$ is exist according to the equations of
motion (2.7). Following the induction process in the same way as in
\cite{15,17}, we find that the $n$-th conserved currents $V_{\pm}^{(n)}$ can be
given in the form
\begin{equation}
\begin{array}{ll}
V_+^{(n)}\,=\,-D_+\chi^{(n)},\\
V_-^{(n)}\,=\,D_-\chi^{(n)}.
\end{array}
\end{equation}
\par
Also, following the procedures given in \cite{15,17}, we can obtain a general
recursion relations which give the infinite number of conserved Susy charges
in the form:
\begin{equation}
\begin{array}{ll}
-D_+\chi^{(n)}\,=\,(-1)^{n-1}2C_+\sum_{j=0}^{n-1}(-1)^j\chi^{(j)},\\
D_-\chi^{(n)}\,=\,2C_-\sum_{j=0}^{n-1}\chi^{(j)},
\end{array}
\end{equation}
where $\chi^{(0)}=1$. To find the $n$-th Susy charge, we use the definition
of $D_\pm$ and from (3.4), it is easy to obtain
\begin{equation}
(\theta_+\partial_{\theta_-}-\theta_-\partial_{\theta_+}-i\theta_-\theta_+
\partial_x)\chi^{(n)}\,=\,
2\theta_+C_-\sum_{j=0}^{n-1}\chi^{(j)}\,+\,(-1)^{n-1}
2\theta_-C_+\sum_{j=0}^{n-1}(-1)^j\chi^{(j)},
\end{equation}
\begin{equation}
(\theta_+\partial_{\theta_-}+\theta_-\partial_{\theta_+}+\theta_-\theta_+
\partial_y)\chi^{(n)}\,=\,
2\theta_+C_-\sum_{j=0}^{n-1}\chi^{(j)}\,-\,(-1)^{n-1}
2\theta_-C_+\sum_{j=0}^{n-1}(-1)^j\chi^{(j)}.
\end{equation}
\par
Writing the supercharge $\chi^{(n)}$ in components as
\begin{equation}
\chi^{(n)}\,=\,\xi_0^{(n)}\,+\,\theta_-\xi_1^{(n)}\,+\,\theta_+\xi_2^{(n)}
\,+\,\theta_-\theta_+\xi_3^{(n)},
\end{equation}
with $\xi_0^{(0)}=1$ and $\xi_i^{(0)}=0;\,i=1,2,3$. Then inserting (3.7) into
(3.5) and (3.6) and solving the results component by component, we obtain
\begin{equation}
\xi_1^{(n)}\,=\,(1-\lambda)\psi_+\sum_{j=0}^{n-1}\xi_0^{(j)},
\end{equation}
\begin{equation}
\xi_2^{(n)}\,=\,(1+\lambda)\psi_-\sum_{j=0}^{n-1}(-1)^{n+j-1}\xi_0^{(j)},
\end{equation}
\begin{equation}
\partial_x\xi_0^{(n)}\,=\,\sum_{j=0}^{n-1}
[J_-\xi_0^{(j)}\,+\,(-1)^{n+j}J_+\xi_0^{(j)}\,+\,
i(1-\lambda)\psi_+\xi_1^{(j)}\,+\,
i(-1)^{n+j-1}(1+\lambda)\psi_-\xi_2^{(j)}],
\end{equation}
\begin{equation}
\partial_y\xi_0^{(n)}\,=\,\sum_{j=0}^{n-1}
[-iJ_-\xi_0^{(j)}\,+\,i(-1)^{n+j}J_+\xi_0^{(j)}\,+\,
(1-\lambda)\psi_+\xi_1^{(j)}\,+\,
(-1)^{n+j}(1+\lambda)\psi_-\xi_2^{(j)}].
\end{equation}
Substituting $\xi_1$ and $\xi_2$ from (3.8) and (3.9) into (3.10) and (3.11), 
we obtain
\begin{equation}
\begin{array}{ll}
\partial_x\xi_0^{(n)}\,=\,\sum_{j=0}^{n-1}
[J_-\xi_0^{(j)}\,+\,(-1)^{n+j}J_+\xi_0^{(j)}]\\
\qquad\qquad\,+\,
\sum_{\ell=0}^{n-2}\sum_{j=0}^\ell
[(1-\lambda)J_-^F\xi_0^{(j)}\,+\,
(-1)^{n+j}(1+\lambda)J_+^F\xi_0^{(j)}],
\end{array}
\end{equation}
\begin{equation}
\begin{array}{ll}
\partial_y\xi_0^{(n)}\,=\,\sum_{j=0}^{n-1}
[-iJ_-\xi_0^{(j)}\,+\,i(-1)^{n+j}J_+\xi_0^{(j)}]\\
\qquad\qquad\,+\,
\sum_{\ell=0}^{n-2}\sum_{j=0}^\ell
[-i(1-\lambda)J_-^F\xi_0^{(j)}\,+\,
i(-1)^{n+j}(1+\lambda)J_+^F\xi_0^{(j)}],
\end{array}
\end{equation}
where $J_\pm$ and $J_\pm^F$ are defined by (2.13). Equations (3.8), (3.9), 
(3.12) and (3.13) give the general recursion relations for the components
of the $n$-th Susy charge for the Susy WZW $\sigma$ model. We observe that,
if we put $\lambda =0$,  these expressions reduce to the corresponding
relations for the Susy $\sigma$ model without the WZW-term \cite{15}.
Moreover, in addition to $\lambda =0$, if the fermionic fields vanish, our
expressions coincide with those obtained for the purely bosonic theory
\cite{17}.
\setcounter{equation}{0}
\section{Integrability on full plane}
To discuss the integrability of the Susy WZW $\sigma$ model on full plane from
the point of view of the conservation of the Susy charges, we first obtain the
explicit form of the first few Susy charges and then show whether they are 
actually conserved or not. Thus, from the above general recursion relations
(3.8), (3.9) and (3.13), we put $n=1$ and find that the first Susy charge
$\chi^{(1)}$ is given by
\begin{equation}
\chi^{(1)}\,=\,\xi_0^{(1)}\,+\,\theta_-\xi_1^{(1)}\,+\,\theta_+\xi_2^{(1)}
\,+\,\theta_-\theta_+\xi_3^{(1)},
\end{equation}
with
\begin{equation}
\begin{array}{lll}
\xi_0^{(1)}\,=\,-i\int_{-\infty}^\infty (J_-\,+\,J_+)dy,\\
\xi_1^{(1)}\,=\,(1\,-\,\lambda)\psi_+,\\
\xi_2^{(1)}\,=\,(1\,+\,\lambda)\psi_-.
\end{array}
\end{equation}
For $n=2$, we see that the second Susy charge $\chi^{(2)}$ is given by
\begin{equation}
\chi^{(2)}\,=\,\xi_0^{(2)}\,+\,\theta_-\xi_1^{(2)}\,+\,\theta_+\xi_2^{(2)}
\,+\,\theta_-\theta_+\xi_3^{(2)},
\end{equation}
with
\begin{equation}
\begin{array}{lll}
\xi_0^{(2)}\,=\,-i\int_{-\infty}^\infty [J_-\,-\,J_+\,+\,J_-^F\,-\,J_+^F 
\,+\,(J_-+J_+)\xi_0^{(1)}\,-\,\lambda(J_-^F+J_+^F)]dy,\\
\xi_1^{(2)}\,=\,(1\,-\,\lambda)\psi_+(1\,+\,\xi_0^{(1)}),\\
\xi_2^{(2)}\,=\,-(1\,+\,\lambda)\psi_-(1\,-\,\xi_0^{(1)}).
\end{array}
\end{equation}
Also, the third Susy charge $\chi^{(3)}$ is
\begin{equation}
\chi^{(3)}\,=\,\xi_0^{(3)}\,+\,\theta_-\xi_1^{(3)}\,+\,\theta_+\xi_2^{(3)}
\,+\,\theta_-\theta_+\xi_3^{(3)},
\end{equation}
with
\begin{equation}
\begin{array}{lll}
\xi_0^{(3)}\,=\,-i\int_{-\infty}^\infty [J_-\,+\,J_+\,+\,2(J_-^F\,+\,J_+^F) 
\,+\,(J_--J_+)\xi_0^{(1)}\,+\,(J_-^F-J_+^F)\xi_0^{(1)}\\
\qquad\qquad 
+\,(J_-+J_+)\xi_0^{(2)}\,-\,2\lambda (J_-^F-J_+^F)\,-\,\lambda (J_-^F+J_+^F)
\xi_0^{(1)} ]dy,\\
\xi_1^{(3)}\,=\,(1\,-\,\lambda)\psi_+(1\,+\,\xi_0^{(1)}\,+\,\xi_0^{(2)}),\\
\xi_2^{(3)}\,=\,(1\,+\,\lambda)\psi_-(1\,-\,\xi_0^{(1)}\,+\,\xi_0^{(2)}),
\end{array}
\end{equation}
and so on.
\par
Next, we show whether these Susy charges, in particular, the Susy Noether
charges $\xi_0^{(n)}$, are conserved on the full plane or not. To see this,
from (4.2) we have
\begin{equation}
\begin{array}{lll}
\frac{d\xi_0^{(1)}}{dx}\,=\,-i\int_{-\infty}^\infty \frac{\partial}{\partial x}
(J_-\,+\,J_+)dy\\
\qquad
=\,-i\int_{-\infty}^\infty \frac{\partial J_x}{\partial x}\,dy\,=\,
i\int_{-\infty}^\infty \frac{\partial J_y}{\partial y}\,dy,
\end{array}
\end{equation}
where we have used (2.14) and (2.15). As
\begin{equation}
\lim_{y\rightarrow \pm\infty}J_x\,=\,
\lim_{y\rightarrow \pm\infty}J_y\,=\,0,
\end{equation}
we see that $d\xi_0^{(1)}/dx =0$. This implies that the first Susy Noether 
charge is conserved on full plane.
\par
Also, for the second Susy Noether charge, we have from (4.4)
\begin{equation}
\frac{d\xi_0^{(2)}}{dx}\,=\,-i\int_{-\infty}^\infty \frac{\partial}{\partial x}
[J_-\,-\,J_+\,+\,J_-^F\,-\,J_+^F\,+\,(J_-+J_+)\xi_0^{(1)}\,-\,    
\lambda (J_-^F+J_+^F)] dy.
\end{equation}
In terms of the $x-y$ variables, we can use (2.15) and (2.16) and find that
\begin{equation}
\frac{d\xi_0^{(2)}}{dx}\,=\,-i\int_{-\infty}^\infty \frac{\partial}{\partial x}
[iJ_y\,+\,iJ_y^F\,+\,J_x\xi_0^{(1)}\,-\,\lambda J_x^F] dy.
\end{equation}
Moreover, if we use (2.8) and (2.14), and the recursion relations (3.12) and
(3.13) for $n=1$, we rewrite this equation as
\begin{equation}
\frac{d\xi_0^{(2)}}{dx}\,=\,\int_{-\infty}^\infty \frac{\partial}{\partial y}
[J_x\,+\,J_x^F\,+\,iJ_y\xi_0^{(1)}\,-\,i\lambda J_y^F] dy.
\end{equation}
Thus from (4.8) and
\begin{equation}
\lim_{y\rightarrow \pm\infty}J_x^F\,=\,
\lim_{y\rightarrow \pm\infty}J_y^F\,=\,0,
\end{equation}
we see that $d\xi_0^{(2)}/dx=0$, that is the second Susy Noether charge is
conserved on full plane.
\par
Similarly, we can use the relations which transform all the quantities
into the $x-y$ variables as before and integrating the results as a total
derivatives then use (4.8) and (4.12) to prove the conservation of all the
Susy Noether charges of the Susy WZW $\sigma$ model on full plane.
At this stage, we see that, as expected, the Susy WZW $\sigma$ model is said 
to be integrable on full plane as a concequence of the conservation of the 
Susy charges.
\setcounter{equation}{0}
\section{Integrability on half plane}
To discuss the integrability of the Susy WZW $\sigma$ model on half plane
with free boundary at $y=0$, $i.e.$, Neumann condition,
we first introduce the model. This model can be defined by the action (2.3)
in which $-\infty < x < \infty$ and $0\le y < \infty$. It is easy to see
that the minimization of this action on half plane leads to the equations
of motion (2.5) and the following free boundary conditions:
\begin{equation}
J_y(x,0)\,=\,0,\qquad\quad \psi_1(x,0)\,=\,\psi_2(x,0)\,=\,0,
\end{equation}
Using these free boundary conditions and the definitions of $J_x$ and $J_y$
given by (2.15) and (2.16), we deduce that
\begin{equation}
J_x(x,0)\,=\,0.
\end{equation}
\par
Let us examine whether the infinite number of nonlocal Susy charges, examined 
before on full plane, are still conserved on half plane or not in the presence 
of Neumann condition. To see this, we examine in particular the conservation
of the first few Susy Noether charges $\xi_0^{(n)}$. From (4.2), the first
Susy Noether charge is written as
\begin{equation}
\xi_0^{(1)}\,=\,-i\int_0^\infty J_x(x,y)\,dy.
\end{equation}
If we proceed as for equation (4.7), we see that
\begin{equation}
\frac{d\xi_0^{(1)}}{dx}\,=\,-iJ_y(x,0),
\end{equation}
which vanishes according to (5.1). Therefore, the first Susy Noether charge
is conserved on half plane as a consequence of the free boundary condition.
\par
For the second Susy Noether charge $\xi_0^{(2)}$, we have from (4.11) that
\begin{equation}
\frac{d\xi_0^{(2)}}{dx}\,=\,\int_{0}^\infty \frac{\partial}{\partial y}
[J_x\,+\,J_x^F\,+\,iJ_y\xi_0^{(1)}\,-\,i\lambda J_y^F] dy.
\end{equation}
Integrating as a total derevatives and using the free boundary conditions 
(5.1) and equations (5.2), (4.8) and (4.12), we deduce that $d\xi_0^{(2)}/dx 
=0$. This means that the second Susy Noether charge is also conserved on half
plane with Neumann condition.
\par
In the same manner, we can prove that all the nonlocal Susy charges obtained
before are conserved on half plane. Therefore, the Susy WZW $\sigma$ model is 
still integrable on half plane with Neumann condition as a concequence of the
conservation of the Susy charges too.
\par
In conclusion, we have studied the classical integrability of the Susy WZW
$\sigma$ model on both full and half plane. We have constructed general
recursion relations for the infinite number of nonlocal Susy charges of the
model. We have proved that these Susy charges are conserved on full plane
and thus the model, as expected, is integrable on full plane. Also, we have
found that these Susy charges are conserved too on half plane with Neumann 
condition. Consequently, the Susy WZW $\sigma$ model on half plane is still 
integrable for Neumann condition.

}

\vfill

\begin{thebibliography}{99}
\bibitem{1}see for example: W. J. Zakrzewski, "Low Dimensional Sigma Models" 
(Adam Hilger, Bristol, 1989).
\bibitem{2}E. Witten, {\it Nucl. Phys.} {\bf B223} (1983) 422.
\bibitem{3}J. Wess and B. Zumino, {\it Phys. Lett.} {\bf 37B} (1971) 95.
\bibitem{4}A. Polyakov and P. Wiegmann, {\it Phys. Lett.} {\bf 131B} (1983) 
121; {\bf 141B} (1984) 223. 
\bibitem{5}E. Witten, {\it Commun. Math. Phys.} {\bf 92} (1984) 455.
\bibitem{6}E. Witten, {\it Phys. Rev.} {\bf D16} (1977) 2991.
\bibitem{7}
B. Piette, R. A. Zait and W. J. Zakrzewski, {\it Zeitschrift fur Physik}
{\bf C44} (1989) 111.
\bibitem{8}E. K. Sklyanin, {\it J. Phys.} {\bf A21} (1988) 2375;\\
S. Ghoshal and A. Zamolodchikov, {\it Int. J. Mod. Phys.} {\bf A9} (1994)
3841;\\
E. Corrigan, P. E. Dorey, R. H. Rietdijk and R. Sasaki, {\it Phys. Lett.}
{\bf B333} (1994) 83;\\
Z. M. Sheng and H. B. Gao, {\it Int. J. Mod. Phys.} {\bf A11} (1996) 4089.
\bibitem{9}A. LeClair, {\it Nucl. Phys.} {\bf B450} (1995) 753;\\
X. G. Wen, {\it Phys. Rev.} {\bf B44} (1991) 5708.
\bibitem{10}E. Corrigan and Z. Sheng, {\it Int. J. Mod. Phys.} {\bf A12} 
(1997) 2825.
\bibitem{11}M. Moriconi and A. De Martino, {\it Phys. Lett.} {\bf B447} 
(1999) 292.
\bibitem{12}W. He and L. Zhao, {\it Phys. Lett.} {\bf B570} (2003) 251. 
\bibitem{13}J. M. Evans, M. Hassan, N. J. MacKay and A. J. Mountain, {\it 
Nucl. Phys.} {\bf B580} (2000) 605.
\bibitem{14}C. Albertsson, U. Lindstr\"om and M. Zabzine, {\it Nucl. Phys.} 
{\bf B678} (2004) 295.
\bibitem{15}M. F. El-Sabbagh and R. A. Zait, {\it Physica Scripta} {\bf 47} 
(1993) 9.
\bibitem{16}E. Brezin, C. Itzykson, J. Zinn-Justin and J.-B. Zuber,
{\it Phys. Lett.} {\bf B82} (1979) 442.
\bibitem{17}M. F. El-Sabbagh and R. A. Zait, "Local and Nonlocal Conserved 
Currents for the Purely Bosonic $U(N)$ $\sigma$ Model", {\it The Fourth 
Conference on Theoretical and Applied Mechanics} (Cairo, Egypt 1991).




\vfil


\end{thebibliography}
\end{document}